\documentclass[usenatbib]{an}

\usepackage{graphicx}
\usepackage{times}
\begin{document}

\Pagespan{1}{}
\Yearpublication{2008}%
\Yearsubmission{2008}%
\Month{8}%
\Volume{???}%
\Issue{???}%

\title{RoboNet-II: \\ Follow-up observations of microlensing events with a robotic network of telescopes}
\author{Y.~Tsapras\inst{1,2} \and  R.~Street\inst{1,13} \and  K.~Horne\inst{3} \and  C.~Snodgrass\inst{6} \and  M.~Dominik\inst{3} \and  A.~Allan\inst{7} \and  I.~Steele\inst{2} \and  D.M.~Bramich\inst{4} \and  E.S.~Saunders\inst{1,7} \and  N. Rattenbury\inst{5} \and  C.~Mottram\inst{2} \and  S.~Fraser\inst{2} \and  N.~Clay\inst{2} \and  M.~Burgdorf\inst{2} \and  M.~Bode\inst{2} \and  T.A.~Lister\inst{1} \and  E.~Hawkins\inst{1} \and  J.P.~Beaulieu\inst{9} \and  P.~Fouqu\'{e}\inst{8} \and  M.~Albrow\inst{10} \and  J.~Menzies\inst{11} \and  A.~Cassan\inst{12} \and D.~Dominis-Prester\inst{14}
}
\titlerunning{RoboNet-II}
\authorrunning{Y. Tsapras et al.}
\institute{
Las Cumbres Observatory Global Telescope network, 6740 Cortona Drive, suite 102, Goleta, CA 93117, USA \and
Astrophysics Research Institute, Liverpool John Moores University, Liverpool CH41 1LD, UK \and
School of Physics and Astronomy, Univ. of St Andrews, Scotland KY16 9SS, UK  \and
Isaac Newton Group of Telescopes, Apartado de Correos 321, E-38700 Santa Cruz de la Palma, Canary Islands, Spain  \and
Jodrell Bank Centre for Astrophysics, The University of Manchester, Manchester, M13 9PL, UK \and
European Southern Observatory, Alonso de Cordova 3107, Vitacura, Santiago, Chile  \and
School of Physics, University of Exeter, Stocker Road, Exeter EX4 4QL, UK  \and
LATT, Universit\'{e} de Toulouse, CNRS, France \and
Institut d'Astrophysique de Paris, Universit\'{e} Pierre et Marie Curie, CNRS UMR7095, 98bis Boulevard Arago, 75014 Paris, France \and
University of Canterbury, Department of Physics and Astronomy, Private bag 4800, Christchurch 8020, New Zealand \and
South African Astronomical Observatory, PO box 9, Observatory 7935, South Africa \and
Astronomisches Rechen-Institut, M\"{o}nchhofstr. 12-14, 69120 Heidelberg, Germany \and 
Dept. of Physics, Broida Hall, University of California, Santa Barbara CA 93106-9530, USA \and
Department of Physics, University of Rijeka, Omladinska 14, 51000 Rijeka, Croatia
}

\received{Aug 2008}
\accepted{??? 2008}
\publonline{later}

\keywords{
Stars: planetary systems, extrasolar planets, microlensing \\
Techniques: photometric, difference image analysis, intelligent agent, robotic telescope network
}

\abstract{%
RoboNet-II uses a global network of robotic telescopes to perform follow-up observations of microlensing events in the Galactic Bulge. The current network consists of three 2m telescopes located in Hawaii and Australia (owned by Las Cumbres Observatory) and the Canary Islands (owned by Liverpool John Moores University).  In future years the network will be expanded by deploying clusters of 1m telescopes in other suitable locations.
A principal scientific aim of the RoboNet-II project is the detection of cool extra-solar planets by the method of gravitational microlensing. 
These detections will provide crucial constraints to models of planetary formation and orbital migration. 
RoboNet-II acts in coordination with the PLANET microlensing follow-up network and uses an optimization algorithm (``web-PLOP'') to
select the targets and a distributed scheduling paradigm (eSTAR) to execute the observations.  Continuous automated assessment of the observations
and anomaly detection is provided by the ARTEMiS system.
}

\maketitle

\section{Introduction}
When a massive stellar object intercepts the line of sight of an observer and a bright background source, the light rays
originating from the source are bent by the gravity of the intervening object. We call this phenomenon gravitational lensing, where the intervening object is termed the {\it lens} and the background object the {\it source}. Depending on the nature of the lens object and relative alignment of observer-lens-source, the lensing effect creates multiple arc-like images of the background source around the edge of the gravitational influence of the lens. If the source, lens and observer are perfectly aligned, the multiple images merge and appear as a bright ring around the lens (Chwolson 1924, Einstein 1936). This ring is usually referred to as the {\it Einstein ring} of the lens and its radius depends on the lensing mass.

Microlensing is a special case of gravitational lensing in which the images of the source are so close to each other that they cannot be independently resolved. In this case, one observes an increase in the brightness of the background source as the lens traverses it and a gradual dimming back to the original source brightness as the lens moves away (Paczynski 1986). This change in brightness plotted versus time is called the {\it event lightcurve}. Planets orbiting the lens may be detectable by the revealing tell-tale signs they leave on the microlensing event lightcurve (Mao \& Paczynski 1991, Paczynski 1991). 

Nowadays, microlensing is one of the methods routinely used to detect extra-solar planets and is particularly sensitive to low-mass planets (down to the mass of the Earth) orbiting a few AU from their host stars (Beaulieu et al. 2006, Bond 2004, Gould et al 2006, Han 2007, Udalski 2005). In this respect, it is complementary to the ongoing transit and radial velocity searches which are more sensitive to giant planets in close orbits (Butler 2006, Cameron et al. 2007, Fischer et al. 2007, Lister at al. 2007).

The remainder of the article is structured in the following way: Section 2 describes the network of telescopes. Section 3 gives an outline of the robotic control system that is used to operate the telescopes in automatic mode. Section 4 presents how the observations are queued and handled by autonomous agents. Our method of data acquisition and processing is discussed in section 5. Finally, section 6 briefly presents how we fit the microlensing events. We conclude with a summary of the paper in section 7.

\section{An expanding network of telescopes}
RoboNet-II continues on the path laid out by the pilot program, RoboNet-1.0, which first used a global network of fully 
robotic 2m telescopes. For Robonet-II in 2008, the same telescope resources are used, while the software that drives, prioritises 
and reduces the observations has been upgraded. The telescopes employed are:
\begin{itemize}
\item The Liverpool Telescope (LT) in La Palma, Canary Islands (owned by the ARI\footnote{Astrophysics Research Institute, Liverpool John Moores University, UK})
\item The Faulkes Telescope North (FTN) in Maui, Hawaii (owned by LCOGT\footnote{Las Cumbres Observatory Global Telescope network, Santa Barbara, California, US})
\item The Faulkes Telescope South (FTS) in Siding Springs, Australia (owned by LCOGT)
\end{itemize}

The telescopes are controlled by a web of interacting programs, which are discussed in more detail in sections 3 to 6. These can function as a single instrument, by optimizing and sharing the load of observations, or as individual instruments which perform the scheduled observations separately. The networked operation allows round-the-clock imaging of astronomical targets. 

The LT (Steele et al. 2004) was designed and built by Telescope Technologies Limited (TTL, now part of LCOGT\footnote{http://www.lcogt.net}) as the prototype of their production-line range of two-metre class telescopes. The telescope itself is a 2m Cassegrain reflector, with Ritchey-Chr\'{e}tien hyperbolic optics, on an alt-azimuth mount. 
A total of five different instruments can be mounted at the Cassegrain focus, one in the `straight through' position and four more on side ports accessible by a rotating `science fold' tertiary mirror. The Faulkes Telescopes are of identical design to the LT.

Current instruments on the LT are two optical cameras (RATCam,RISE), an optical polarimeter (RINGO), and an infrared imaging array (SupIRCam)\footnote{http://telescope.livjm.ac.uk/Info/TelInst/Inst/content.html}. The Faulkes telescopes are equipped with an E2V-4240 2k$\times$2k optical CCD camera (MEROPE) and a reserve Apogee Alta-U camera (HawkCAM) with plans to replace these with Spectral Instruments 600 series cameras using Fairchild Imaging CCD486 devices.

The network will be gradually expanded (Hidas et al. 2008) with the introduction of 18 new 1m and 24 0.4m telescopes, all of which are expected to be fully integrated and in operation by 2011 at distributed sites around the world. 

\section{The Robotic Control System}
Each networked telescope is controlled by a Robotic Control System (RCS) (Fraser \& Steele 2004). This system runs on a computer at the telescope site along with the Telescope Control System (TCS) and Instrument Control Systems (ICS). The RCS instructs the software controlling the telescope and various instruments and is responsible for beginning and end-of-night operations as well as carrying out the observations. Calibration images are automatically taken and the preprocessed images are transferred to the archives.

The RCS communicates with the TCS to control the telescope functions (slewing, tracking, autoguiding, acquisition, enclosure opening/closing etc). The whole system operates intelligently whereby if something goes wrong such as an instrument failing to initialize properly, it attempts to fix itself (Mottram 2006). Additionally, the telescopes have local weather monitoring systems which feed information on humidity, cloud cover, precipitation, wind speed and temperature to the RCS and lower level control systems. If any of these parameters exceed their allowed ranges (or if any of the sensors fail), the enclosure is automatically closed.

When atmospheric conditions (seeing, extinction) are too poor to allow normal science programs to run, the RCS switches the telescope to a background observing mode. In this operating mode, a series of photometric standard stars are observed regularly in order to monitor changes in atmospheric conditions. Normal operation may be restored when the atmospheric parameters return within acceptable limits.

In terms of scientific observations, the RCS takes into account input from the phase II database (see section 4.1), which is stored locally at each telescope site. This is where all the observing programs with their specifications are stored. The information is used by the RCS to determine the control operations required for the telescope and instrument systems. A separate database is kept at all networked telescopes and all of them can be queried externally. 

Intelligent Agents (discussed in section 4.3) schedule the submission of observation requests to the individual telescopes in order to ensure that time-critical phenomena are promptly and automatically observed by the telescope that is best suited to make the observations.

The data generated by these observations are then fed through reduction pipelines where preliminary quality assessments are made and the information is then passed back to the intelligent agent so it can adjust its reaction accordingly. For example, if an observation has failed, it will decide whether it should be re-submitted or moved to another telescope.

The RCS accepts input from two sources. The first one is the Observer Support System (OSS). It controls access to the database of observations uploaded by the users to the telescope (via a GUI) or submitted automatically by external agents. These external intelligent agents can modify or even add new observations to the existing database. The OSS also provides the observation scheduler. The second source of input is from the Target of Opportunity Control System (TOCS). This is an override program which can request the RCS to interrupt the current observing schedule and initiate immediate observations of a priority target. The TOCS can be started automatically via external triggers or may be invoked manually with any instrument and filter combination required. Typical examples of this are Gamma-Ray bursts (Guidorzi et al. 2006) and caustic crossing entries in microlensing events. 

Transitions between the different RCS operational states can take a few seconds or may last a few minutes, depending on the states involved. Offsets between the various instruments and filters are predefined in the configuration files of the RCS.

\section{Scheduling and obtaining observations}
\subsection{Submission of observing requests}
The Phase II database contains details of the observing requests (targets, configuration, exposures) including constraints on timing and permitted observing conditions. The database also contains details of resource accounting (time used by programs) and a history of observations performed.  
\begin{figure}
\begin{centering}
\includegraphics[width=82mm]{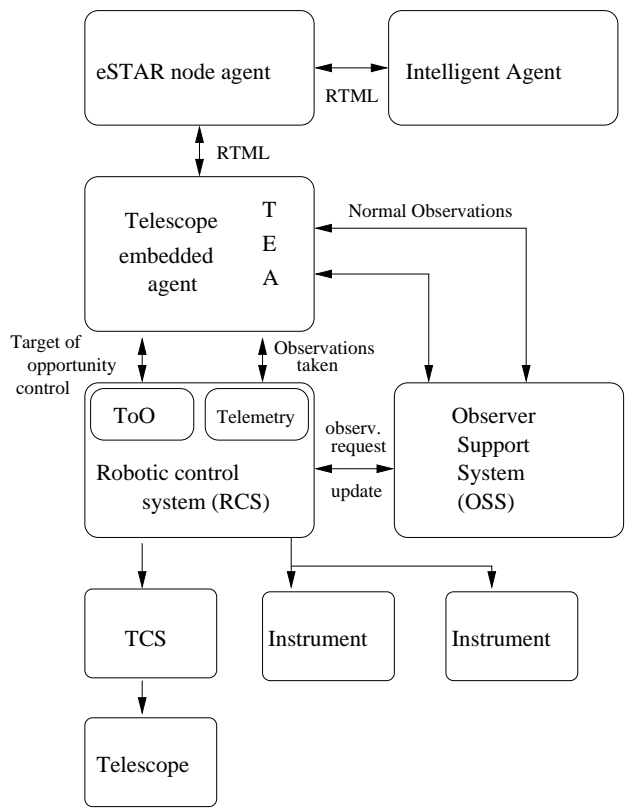}.
\caption[TEA architecture and dependencies]{The Telescope Embedded Agent (TEA) architecture and its dependencies. Automatic external "Intelligent Agents" (IA) communicate with site specific (local) embedded agents and request observations from the Robotic Control System (RCS) of the telescope. The opportunity to bypass normal observations is also available by issuing a Target of Opportunity (ToO) override. Such overrides can be requested manually or automatically.}
\label{fig:tea}
\end{centering}
\end{figure}

Each observation, regardless of whether it was submitted manually or automatically, is part of a submitted {\it proposal}. Proposals have an activation and expiry date as well as a fixed time allowance and time allocations under various observing conditions. Each proposal is also given a scientific priority level which is decided by the Time Allocation Committees (TAC).

Individual {\it proposals} contain many {\it groups} of observations. A {\it group} is a collection of observations which must be scheduled and performed as a unit. To submit a {\it group}, the user has to specify a number of parameters: The activation and expiration dates of the proposal, a limit on the maximum allowable sky brightness, the worst allowable seeing conditions and a monitoring interval (for periodic observations).

Each {\it group} consists of individual {\it observation} requests, each with a specific exposure time and repeat count, target position on the sky and instrument selection and configuration. The requested observations are processed by the scheduler before being executed as discussed in the previous section.

There are five types of observing groups: i) `{\it{Flexible}}' groups can be scheduled at any time and are typically  one-off observations of targets. ii) `{\it{Fixed}}' groups are also one-off observations but can only be performed at a very specific time and prevent other groups from being considered for scheduling. iii) `{\it{Monitoring}}' groups are periodic observations of the same target at a fixed interval. iv) `{\it{Ephemeris}}' groups specify observations that should be obtained at a given phase in a variable objects cycle, although there are no restrictions on which cycle these should be obtained on. v) `{\it{MinInterval}}' groups are performed {\it{at least}} the specified interval apart.

Microlensing observation groups are typically scheduled as {\it{Monitoring}} requests.

\subsection{The scheduler}
When the RCS is ready to perform an observation, all active groups in the database are sorted by the real-time scheduling algorithm which picks the next `best' observation to perform by optimizing against a number of selection/scoring criteria (Fraser 2006). 

The scheduler operates in a rapidly changing environment. Poor weather, changes in observing conditions, unexpected mechanical faults or software glitches in addition to the overrides by other software agents (ToO) can occur over short timescales. This would make any long-term scheduling decisions brittle. 

The scheduler therefore uses a simple dispatch mechanism which selects just a single group of observations to perform at each invocation. This has the advantage of always selecting observations best matched to the {\it{current}} conditions (local optimization) but has the disadvantage that, since no look-ahead is performed to check the effects on future observing possibilities, global optimization criteria are not maximized. 
 
There are plans to incorporate a degree of forward planning in future releases of the Scheduler (Fraser \& Steele 2008, Saunders et al. 2008).

\subsection{Intelligent agents}
An intelligent agent is a program that can make autonomous decisions. Within the RoboNet context, it is used to mine the on-line catalogues and databases of observations and tell each telescope on the network (via the telescope embedded agent which talks directly to the telescope scheduler) what to observe and when (Allan et al. 2004, 2006). 

The observations are requested using the method of {\it{adaptive dataset planning}} (Naylor et al. 2004), by which information from the latest analysis of the available data so far are folded into the process of developing the plan for the next series of observations to be queued. 

The agent is designed to be both proactive and responsive. In the microlensing case, this means that it must react and request observations in the guise of a target of opportunity (ToO) as soon as it is alerted to an ongoing anomaly. The intelligent agent software was developed by the eSTAR group\footnote{http://www.estar.org.uk/} and is currently hosted at Exeter. 

\subsection{Telescope embedded agent and RTML}
\begin{figure*}
\begin{centering}
\includegraphics[width=160mm]{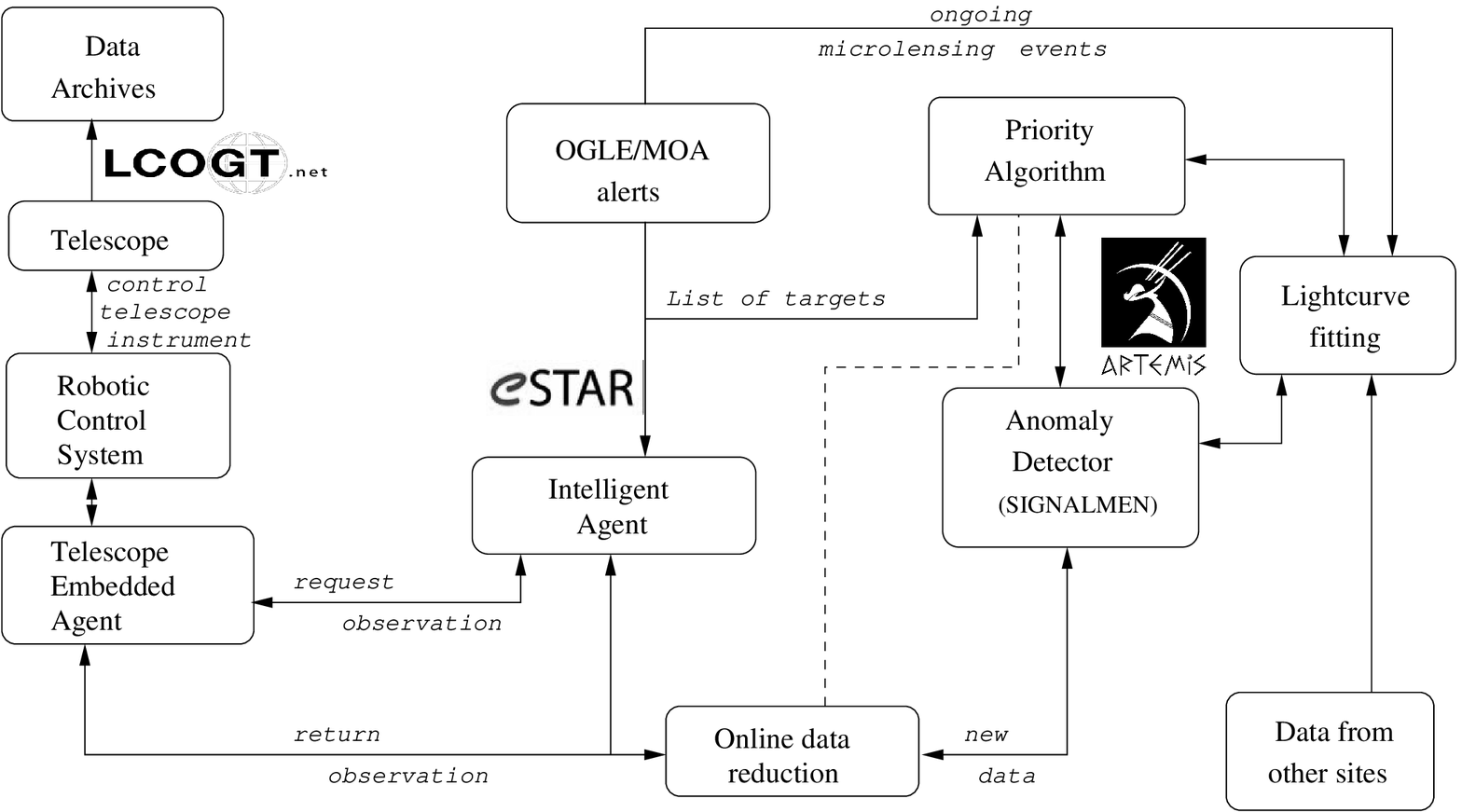}.
\caption[The communication network between the various systems.]{RoboNet-II microlensing follow-up system architecture. A priority algorithm selects the list of microlensing events to follow-up on at any given time. The list is imported by the eSTAR intelligent agent and submitted as observing requests to the telescope embedded agents. Once an observation of a microlensing event is performed, the image is immediately transferred to the data processing centre and the image reduction pipeline is initiated. Lightcurve data are made promtly available and the anomaly detector identifies suspected anomalies in real time.}
\label{fig:process}
\end{centering}
\end{figure*}
A telescope embedded agent (TEA) handles the requests made by the various intelligent agents (each representing a different science program) and updates the observing database. It is also authorized to make target of opportunity observations in real time by overriding the normal observing program. 

Intelligent agents, communicating with the TEA at each telescope, receive information about how suitable that telescope is for performing the requested observation. Based on this information, they pick the telescope best suited to perform the observation and send a request prompting that TEA to schedule it. 

The intelligent agents and the TEAs communicate by RTML (Remote Telescope Markup Language) (Hessman 2006). 
This provides a means of making a telescope independant description of an observing request using 
an XML language defintion.  It is used for observation {\it scoring} and {\it requesting} (Mottram 2006) and allows the TEAs to send reduced data products back to the intelligent agents for analysis. RTML has become the standard for interface communication of robotic telescope networks and can allow inter-operation between different networks.

A flowchart depicting how the various systems communicate with each other is shown in figure~\ref{fig:tea}.

\subsection{Data transfer and archiving}
Once the observations have been performed, the images are transferred every 10 minutes from each telescope to the archives and are available on a quick-look page so that, if required, an assessment of data quality can be made immediately. When all data for a particular night have been transferred from the telescope sites, they are debiased and flat fielded, using the latest calibration images, and stored in the archives under the appropriate proposal name.

\section{RoboNet-II microlens planet search}
The system described in the previous sections is used by the RoboNet-II microlens program to detect extrasolar planets. We now turn to how the processes we have set up communicate with the eSTAR intelligent agents and how the data acquisition loop is closed.

The OGLE\footnote{http://www.astrouw.edu.pl/$\sim$ogle/} and MOA\footnote{http://www.phys.canterbury.ac.nz/moa/} survey teams now routinely discover $\sim$1000 microlensing events per year. When a new event is discovered, an immediate alert is issued on the internet. Planet-hunting follow-up teams 
then decide whether the event is promising and warrants follow-up observations.

With dozens of events going on at the same time, not all of them can be observed with the required sampling frequency. So observers are faced with a multifaceted dilemma: which events should be observed, how long for and how frequently should each target be observed?
There are two general approaches to these questions that define the observing strategies of the follow-up teams. High magnification events are usually selected as promising targets. For such events, the probability of detection of a Jupiter-like planet in the lensing zone (0.6-1.6 $\theta_{E}$, where $\theta_{E}$ is the angular Einstein ring radius,) approaches 100\% (Griest \& Safizadeh 1998). 

Given that one knows a-priori that the planetary deviation is to occur near the peak of the light curve, a suitable simple strategy is to try to monitor all events that rise towards a high peak quasi-continuously in the peak region with the aim of fully {\it characterizing} the event.

Another approach relies on rapid data reduction, an efficient anomaly detector and the ability of a robotic telescope network to immediately respond to observing override requests. This strategy attempts to maximize the planet {\it discovery rate} (Horne 2008) while the possibility of automated fast responses can lead to the {\it characterization} of the observed events. 

Every data point on a microlensing lightcurve carries information about the presence, or absence, of a planet on either of the two image positions around the Einstein ring\footnote{In the case of a planetary deviation, with sufficient sampling it is possible to distinguish which of the two images the planet is perturbing.}. 

The areas of the detection/exclusion zones associated with each data point increase with magnification and decrease when the photometric uncertainties become larger. The idea here is that events are sampled in such a way so as not to produce overlaps of the detection zones. This translates into an optimal sampling strategy for each event which also allows for the surveying of more events where anomalies may manifest. As soon as such an anomalous data point is detected, the monitoring strategy is immediately changed\footnote{Hence the need for a fully robotic operation} and all available resources are used to {\it characterize} the anomaly. The efficiency of this method will increase with the number of networked telescopes and minimization of the response time.

While RoboNet-II schedules observations according to an optimal sampling scheme in order to maximize the scientific output of the campaign, the few easy targets reaching high peak magnifications are not missed either, and at any time manual overrides can be done. 

Microlensing planet searches have yielded important results in the last few years.
The source passing close to the lens allowed the double catch of a Jupiter/Saturn analogue orbiting
OGLE-2006-BLG-109L (Gaudi et al. 2008a), while the presence of a Neptune-class planet was inferred from observations of event OGLE-2006-BLG-169 (Gould et al. 2006). Furthermore, a planet several times
more massive than Jupiter was revealed from monitoring OGLE-2005-BLG-071 (Udalski et al. 2005). 

However, an observed off-peak planetary deviation constitutes the most exciting discovery so far, namely that of OGLE-2005-BLG-390Lb, the first reported icy exoplanet, about 5 times more massive than Earth, and the most Earth-like planet orbiting a star other than the Sun at the time of its announcement (Beaulieu et al. 2006).
All these discoveries involved Robonet-1.0 data.

\subsection{A prioritising algorithm}
The web-PLOP (Planet Lens OPtimistation) software (Snodgrass et al. 2008) was originally developed to compile an optimal list of targets to observe for the RoboNet-1.0 project (Burgdorf et al. 2007), and moreover to provide an easily accessible web-interface to monitor the progress of the observations. Since then it has expanded to accommodate any observing site that wants to use it to select microlensing targets and to keep track of its observations.

For any telescope, a unique profile can be entered and stored, including the telescope characteristics (like effective area, slew time, readout time, etc.), observing conditions (like e.g. sky level), and the total available observing time. 

The optimisation software is part of a top-level system which keeps an up-to-date record of all data from OGLE, MOA, PLANET and RoboNet-II observations, while new point-source point-lens fits are produced whenever new data become available. It uses these to predict the future magnifications of events, and selects those to observe which maximize the planet detection probability for a given telescope and time (Horne 2008).

A direct feedback to web-PLOP from the intelligent agents that communicate with the RoboNet telescopes is provided in the form of the time and observing condition for each data point, allowing a re-prioritisation in real time, adapting for changing conditions at the telescopes, even before the respective photometry becomes available.

The online public data archives are searched in real time by a `detector' which has the ability to issue alerts on suspected anomalies. We describe this next.

\subsection{Automated anomaly detection: SIGNALMEN}
Using an automated anomaly detector (Dominik et al. 2007) allows the discovery of low mass planets without initial high-cadence sampling.  
This permits RoboNet-II to monitor enough events in order to have a fair chance of detecting an Earth-mass planet within the next few years. The anomaly detector exploits the possibility of fast response and flexible scheduling that is provided by the robotic telescope network. Therefore, we can adopt a rather low threshold on the first suspicion of an anomaly by forcing the telescopes to observe the target again if data appear to deviate by an amount exceeding that of 95\% of the previously observed points. Further data are then obtained until a decision about whether an anomaly is present or not can be taken with the desired significance. This decision is taken automatically by the software with no manual intervention.

The alerting system takes into account RoboNet-II real-time photometry but also includes data from other campaigns, PLANET\footnote{http://planet.iap.fr/}, OGLE, MOA, and MicroFUN\footnote{http://www.astronomy.ohio-state.edu/$\sim$microfun/}, as soon as these are released.
Figure~\ref{fig:process} shows the general architecture of the follow-up system.

\subsection{Incoming RoboNet-II microlensing data}
All new RoboNet-II microlensing data are automatically transferred to the project computer server at LCOGT where the quality of each image is checked and a log file of nightly observations is created and stored locally. Any images showing serious defects or where the stellar profile is grossly distorted are tagged and excluded from the analysis. This stage is automated but on a few occasions there have been images that have had to be tagged manually. These images contained a low number of stars and had sky backgrounds with a gradient across the field. Similarly, if a satellite trail or a bad pixel column passes close to or crosses over the target, then such an image is not included in the subsequent analysis. 

With potentially hundreds of images coming in every night, and keeping in mind the future expansion of the network, there is ongoing work to further automate this process.

\subsection{The reduction pipeline}
The data reduction pipeline is running in two modes which we call the {\it Real Time} and the {\it Offline} modes. Both are run at the LCOGT central processing centre in Santa Barbara, California. The {\it Real Time} mode is activated every time a new image of a microlensing target is obtained. To produce a lightcurve, the pipeline requires that a reference image already exists that it can use and that the event has already been identified in that reference image. The software automatically rejects defective images, selects the best available image to use as a reference and identifies the target star using the WCS information available in the image headers.

The {\it Offline} mode is interactive and can be run manually at the central processing centre. Here, image quality can be double checked. Events can be re-analyzed taking into account any new data or by creating new refrerence images to use.

\subsection{Difference image analysis}
At the central processing centre, the data are automatically sorted as they arrive from the telescopes. There are separate directories for every monitored event, filter and telescope combination. The next stage of the processing initiates the Difference Image Analysis (DIA) pipeline built from the Bramich et al. (2005) DIA software which, in turn, is based on the Alard and Lupton DIA methodology (Alard \& Lupton 1998). When initiated, the pipeline checks every catalogued directory for any new obesrvations. If an event has not been observed previously, it creates a reference image using the single best-seeing image. If there are previous observations of an event, but at least one of the new images has better seeing than the current reference image, then the reference image is recreated using the new best-seeing image, and all the event data are reanalyzed.

\begin{figure}
\begin{centering}
\includegraphics[width=82mm]{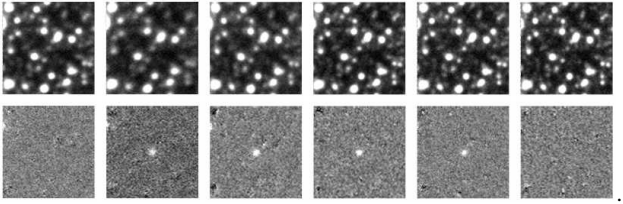}.
\caption[Image subtraction]{Typical results from the image subtraction pipeline. The target star, OGLE-2006-BLG-341S, undergoing microlensing is at the centre of the 50$\times$50 pixel stamp. The top row shows the original images while the bottom row presents the subtracted images where each image from the top row has been aligned to and subtracted from a common reference image. Only stars showing variations in brightness are visible on the subtracted images and the microlensing event stands out clearly.}
\label{fig:target}
\end{centering}
\end{figure}

Once the reference image has been created, every image is geometrically aligned to it. The seeing of the reference image is then degraded to match that of each individual image, and, after photometrically scaling the blurred reference image to match the current image, they are subsequently subtracted (see figure~\ref{fig:target}). Any regions in the resulting difference image that correspond to variable sources (or image defects or saturated stars that have not been perfectly masked) will leave a residual differential flux, which we measure using optimal PSF scaling.

The output files are standardized to a common format that contains the target lightcurve information, the derived trend values used by the pipeline and the photometric conditions under which the data was obtained. This information is then made available via the web-PLOP and ARTEMIS (Dominik et al 2008) webpages. 

We are currently developing a new DIA pipeline. This will feature the latest in resampling algorithms based on the method of cubic O-MOMS (Blu et al. 2001). We are also working on improving the star matching algorithm of Valdes et al. (1995) to make it more robust and produce less scatter in the residuals of the fitted transformation between images. Finally we are working on improving the kernel solution at the heart of the difference imaging technique, using a novel approach to modelling the kernel as a pixel array (Bramich 2008).

\section{Lightcurve fitting}
\begin{figure}
\begin{centering}
\includegraphics[angle=0,width=82mm]{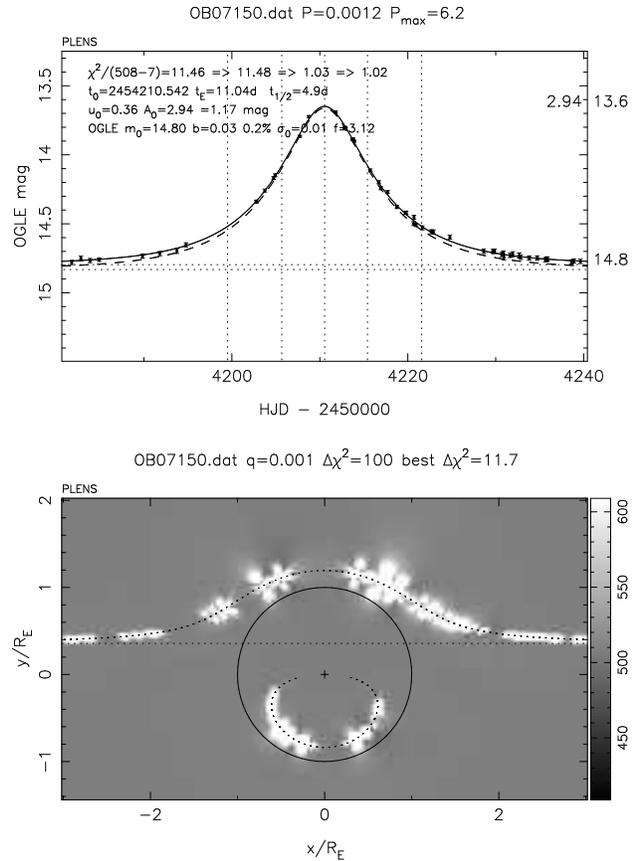}
\caption[Magnitude and delta chi plot for OGLE-2007-BLG-150. Only OGLE data is presented.]{Top panel: Maximum likelihood fits to the data for event OB07150. Only OGLE data are presented to better illustrate the detection zones. The best fitting parameters are shown on the top left corner of the plots. Bottom panel: $\Delta\chi^2$ detection threshold plot for this event. White regions correspond to planet exclusion zones i.e. the data exclude the presence of a planet in those regions.}
\label{fig:plots1}
\end{centering}
\end{figure}
We fit each lightcurve by a global $\chi^2$ minimization over all the available datasets. The parameters of the fit are the standard four describing the shape of a point-source point-lens lightcurve (time of maximum magnification, $t_0$, event timescale, $t_E$ , maximum magnification $A_0$, baseline magnitude, $I_0$) and the two describing the blending flux and magnitude offset of each telescope/filter combination. In addition, we allow for the possibility of rescaling the error bars (an extra 2 parameters), in which case we use a maximum likelihood criterion for the fits as described in Tsapras et al. 2003. The algorithm is set up so that we can adjust either the PSPL parameters only (4), the blend flux \& magnitude offset (6) or allow for rescaling of the error bars (6 or 8 depending on whether a blending fit is required).

For each event, we calculate $\Delta\chi^2$ detection maps. These show the change in $\chi^2$ for a fit with a planet of fixed mass ratio $q$ at position $x,y$ (measured in units of the Einstein ring radius $R_E$) relative to the no planet model (Snodgrass et al. 2004). We calculate these maps by setting up a fine grid of planet positions in $x,y$ on the lens plane and for each of these positions we fit the binary model to the data optimizing all other parameters. It is important to keep the sampling density in $x,y$ dense enough so that no planetary fits are missed. We use a grid-search step-size of $\sqrt{q}/4$. This sets up a fine grid for each selected mass ratio.

Black zones identify regions where the $\Delta\chi^2$ values are above the threshold of detection whereas white zones are the regions where the presence of the planet can be excluded given the data. The grey zones identify all the possible positions where the presence of the planet does not perturb the lightcurve i.e. $\Delta\chi^2$=0. As an illustrative example, we show the lightcurve for the microlensing event OB07150 and the associated $\Delta\chi^2$ map in figures \ref{fig:plots1}(a,b).

The lightcurve fits and $\Delta\chi^2$ maps are generated both for the {\it Real Time} and the {\it Offline modes} of the pipeline. These plots are immediately available online on the project website as soon as they are generated. The results of the fits are fed back to the telescope pipelines and the anomaly detector and if a deviation is found, a ToO override observation is triggered. ToO observations may also be submitted manually.

\section{Summary}
Robonet-II has developed a complete architecure and supporting software to implement that architecture for the automated detection and characterization of exoplanets detected via the microlensing technqiue.  The RoboNet-1.0 pilot programme in previous seasons has returned promising results and contributed to almost all of the microlensing planetary discoveries to date. Current efforts are directed in further automation and restructuring of the scheduling, data acquisition and image processing, improving the alerting system and responses, as well as significant upgrades to the telescope engineering. These involve the deployment of new instruments and electrical safety and performance reliability upgrades. 

LCOGT is in the process of expanding the robotic network of telescopes. Current plans are for 18 new 1m and 24 0.4m telescopes which are expected to be fully integrated and in operation by 2011. The microlensing search for planets, and in particular the method pioneered by the RoboNet project which can potentially make full use of the facilities in an automated way, is a science objective that can be efficiently realized with this network.

\end{document}